\documentclass[preprint,showpacs,preprintnumbers,amsmath,amssymb, prb]{revtex4}
\usepackage{graphicx}
\usepackage{dcolumn}
\usepackage{bm}
\begin{document}
\title{Magnetic properties and complex magnetic phase diagram in non centrosymmetric EuRhGe$_3$ and EuIrGe$_3$ single crystals}
\author{Arvind Maurya}
\affiliation{Department of Condensed Matter Physics and Materials
Science, Tata Institute of Fundamental Research, Homi Bhabha Road,
Colaba, Mumbai 400 005, India}
\author{P. Bonville}
\affiliation{CEA, Centre d'Etudes de Saclay, DSM/IRAMIS/Service de Physique de I'Etat Condens\'e, 91191 Gif-sur-Yvette, France}
\author{R. Kulkarni}
\affiliation{Department of Condensed Matter Physics and Materials
Science, Tata Institute of Fundamental Research, Homi Bhabha Road,
Colaba, Mumbai 400 005, India}
\author{A. Thamizhavel}
\affiliation{Department of Condensed Matter Physics and Materials
Science, Tata Institute of Fundamental Research, Homi Bhabha Road,
Colaba, Mumbai 400 005, India}
\author{S. K. Dhar}
\email{sudesh@tifr.res.in}
\affiliation{Department of Condensed Matter Physics and Materials
Science, Tata Institute of Fundamental Research, Homi Bhabha Road,
Colaba, Mumbai 400 005, India}
\date{\today}
%
\begin{abstract}
We report the magnetic properties of two Eu based compounds, single crystalline EuIrGe$_3$ and EuRhGe$_3$, inferred from magnetisation, electrical transport, heat capacity and $^{151}$Eu M\"{o}ssbauer spectroscopy. These previously known compounds crystallise in the non-centrosymmetric, tetragonal, $I4mm$, BaNiSn$_3$-type structure. Single crystals of EuIrGe$_3$ and EuRhGe$_3$ were grown using high temperature solution growth method using In as flux. EuIrGe$_3$ exhibits two magnetic orderings at $T_{\rm N1}$~=~12.4~K, and $T_{\rm N2}$~=~7.3~K.  On the other hand EuRhGe$_3$ presents a single magnetic transition with a $T_{\rm N}$~=~12~K. $^{151}$Eu M\"{o}ssbauer spectra present evidence for a cascade of transitions from paramagnetic to incommensurate amplitude modulated followed by an equal moment antiferromagnetic phase at lower temperatures in EuIrGe$_3$, the transitions having a substantial first order character. On the other hand the $^{151}$Eu M\"{o}ssbauer spectra at 4.2 and 9~K in EuRhGe$_3$ present evidence of a single magnetic transition. In both compounds a superzone gap is observed for the current density $J\parallel$~[001], which enhances with transverse magnetic field. The magnetisation measured up to 14~T shows the occurrence of field induced transitions, which are well documented in the magnetotransport data as well.  The magnetic phase diagram constructed from these data is complex, revealing the presence of many phases in the $H-T$ phase space.
\end{abstract}

%
\maketitle
\section{Introduction}
The magnetic properties of several Eu-based compounds with composition Eu$TX_3$, where $T$ is a $d$-block transition element and $X$~=~Si or Ge, have been reported in literature~\cite{Neeraj_EuPtSi3,Neeraj_EuPtGe3,Kaczorowski_EuPdGe3,Goetsch_EuNiGe3,Arvind_EuNiGe3,Oleksandr_EuTGe3}. These compounds crystallize in the non-centrosymmetric (NCS) BaNiSn$_3$ -type structure. A transition from paramagnetic to an  incommensurate, amplitude modulated antiferromagnetic state followed by a second transition to a full-moment, antiferromagnetic configuration in  EuPtSi$_3$~\cite{Neeraj_EuPtSi3} and EuNiGe$_3$~\cite{Arvind_EuNiGe3} was inferred from heat capacity and $^{151}$Eu M\"{o}ssbauer spectroscopy. Magnetisation data on a single crystal of EuPtSi$_3$ showed the presence of anisotropy, probably of both crystalline and exchange origin. Only one magnetic transition is observed in EuPtGe$_3$ and EuPdGe$_3$ ~\cite{Neeraj_EuPtGe3,Kaczorowski_EuPdGe3}.  Further, magnetisation data on a single crystal of EuNiGe$_3$ showed that while the $ab$-plane is the hard plane, an unusual staircase-like behavior of magnetisation is observed along the $c$-axis~\cite{Arvind_EuNiGe3}. Thus, a variety of interesting magnetic behaviour is observed in Eu$TX_3$ compounds. The magnetisation in the antiferromagnetic state showing a varying degree of anisotropy which is a priori surprising for a spin-only ($S$~=~7/2; $L$~=~0) ion like Eu$^{2+}$. 

	The existence of iso-structural EuIrGe$_3$ and EuRhGe$_3$ is known and it was of interest to study the magnetic behaviour of these two compounds. We have probed the detailed magnetic properties of single crystalline EuIrGe$_3$ and EuRhGe$_3$ by magnetisation, resistivity and  heat capacity in zero and applied fields, and $^{151}$Eu M\"{o}ssbauer spectroscopy. While this work was in progress, the magnetic properties of single crystalline EuIrGe$_3$ and EuRhGe$_3$ have been reported in literature based on susceptibility measured in a field of 0.1~T, zero-field electrical resistivity and heat capacity~\cite{Oleksandr_EuTGe3}. Our main observations are in agreement with the results reported in ref.~\onlinecite{Oleksandr_EuTGe3}; however, our more extensive data include isothermal magnetisation at selected temperatures, susceptibility measured at a number of applied fields, magnetoresistivity, construction of magnetic phase diagram, observation of the superzone gap at the antiferromagnetic transition and $^{151}$Eu M\"{o}ssbauer spectra. In addition, we also prepared LaRhGe$_3$ and LaIrGe$_3$ as non-magnetic reference compounds and measured their heat capacity and electrical resistivity. 
\section{Experimental}
\label{Exp}
Polycrystalline samples of EuIrGe$_3$ and EuRhGe$_3$ were first prepared by melting Eu (99.9~\%  purity), Ir/Rh (99.99~\%) and Ge (99.999~\%) in an arc furnace under an inert atmosphere of argon. The alloy buttons were flipped over three times and re-melted to ensure proper homogenization. An excess of about 10~\% over the stoichiometric amount was taken for Eu, to compensate the weight loss due to the evaporation of Eu.  Initially, we attempted to grow the single crystals of these compounds by using Sn as a solvent as that choice had proved successful for EuPtSi$_3$ and EuPtGe$_3$, but it did not give desired results. In the second attempt, charges of EuIrGe$_3$ and EuRhGe$_3$ and In (as solvent) were taken in the weight ratio 1:8; placed together in  separate alumina crucibles and sealed in quartz ampoules under a partial pressure of 10$^{-6}$ mbar. The sealed crucibles were placed in a box type resistive heating furnace and heated to $1100\,^{\circ}{\rm C}$ at a rate of $50\,^{\circ}{\rm C}$/hour. After a soaking period of 24~hours, a cooling rate of $2\,^{\circ}{\rm C}$/hour was employed down to $600\,^{\circ}{\rm C}$. The cooling rate was increased to $60\,^{\circ}{\rm C}$/hour below $600\,^{\circ}{\rm C}$.  The single crystals of EuIrGe$_3$ and EuRhGe$_3$ were separated from In-flux by centrifugation. Small traces of indium were washed away by etching the grown crystals in dilute hydrochloric acid.  Polycrystalline samples of non-magnetic reference LaIrGe$_3$ and LaRhGe$_3$ were prepared by the standard technique of arc melting as described above. The magnetisation as a function of field (up to 14~T) and temperature (1.8 to 300~K) was measured using Quantum Design Magnetic properties measurement system (MPMS) and Vibration sample magnetometers (VSM). The electrical resistivity between 1.8 and 300~K and the heat capacity in zero and applied fields was measured in a Quantum Design Physical properties measurement system (PPMS) unit. $^{151}$Eu M\"{o}ssbauer spectra were recorded at various temperatures using a constant acceleration spectrometer with a $^{151}$SmF$_3$ source.  Laue diffraction patterns were recorded on a Huber Laue diffractometer  fitted with an image plate, while powder-diffraction spectra were recorded on a Philips PANalytical set up using Cu-K$\alpha$ radiation.  The crystals were cut by spark erosion electric discharge machine and oriented along the desired planes using a triple axis goniometer and Laue  diffraction  in the back reflection mode. 

\section{Results and Discussion}
\subsection{Structure}
Well faceted crystals having a platelet geometry and typical dimensions of $\sim$~5mm~x~5mm~x~1mm were obtained after centrifuging out the In solvent. The composition of the crystals was confirmed using electron dispersive analysis by x-rays (EDAX). The powder x-ray diffraction spectra, obtained by crushing a few single crystals to powder, could be indexed to tetragonal BaNiSn$_3$~\--type structure.
%
\begin{table}[h]
\centering
\caption{\label{lattice constants} Lattice parameters $a$ and $c$, and unit cell volume $V$ of EuRhGe$_3$ and EuIrGe$_3$ obtained from the Rietveld refinement of x-ray powder diffraction pattern.}
\begin{tabular}{lccc}
\hline\hline
                 & $a$ & $c$ & $V$  \\
                 & (\AA)     & (\AA)     & (\AA$^3$)  \\
\hline
EuRhGe$_3$ & 4.407(3)  & 10.068(7)   & 195.57(7)   \\ 
EuIrGe$_3$ & 4.430(0)  & 10.041(6)   & 197.06(5)   \\       
\hline       
\end{tabular}
\end{table}
%
 The lattice parameters obtained by the Rietveld analysis of the powder diffraction spectra using FullProf software package~\cite{Carvajal} are listed in Table~\ref{lattice constants}; and are in good agreement with the previously reported values~\cite{Oleksandr_EuTGe3, Venturini}.
\subsection{Magnetisation}
\begin{figure}[h]
\includegraphics[width=0.50\textwidth]{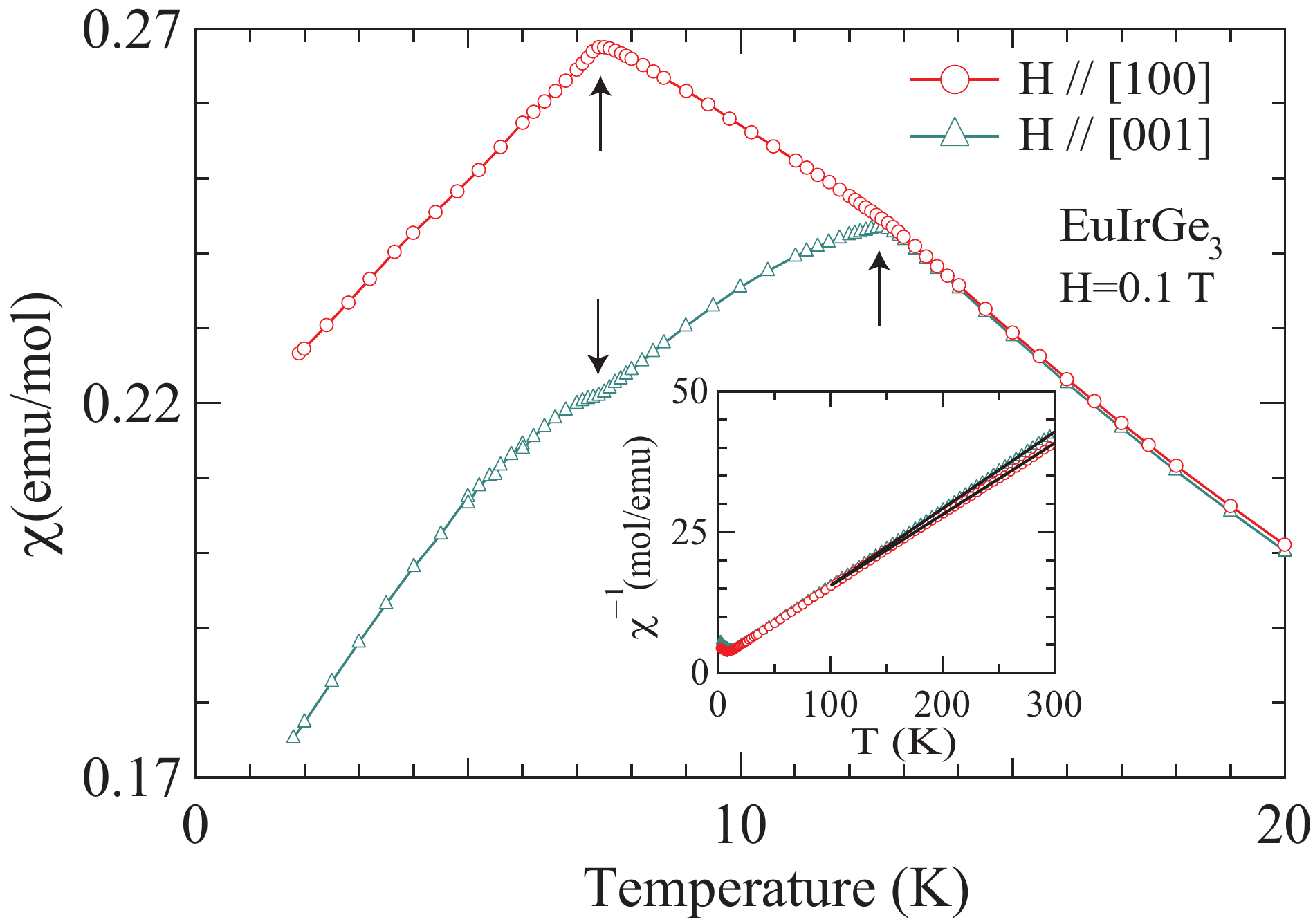}
\caption{\label{EIG3_MT} Magnetic susceptibility and inverse magnetic susceptibility (in inset) of EuIrGe$_3$ at field 0.1~T along [100] and [001].}
\end{figure}

The inverse susceptibility, $\chi^{-1}$, of EuIrGe$_3$, between 1.8 and 300~K, with the magnetic field (0.1~T) applied parallel to [100] and [001] directions, respectively, is shown in the inset of Fig.~\ref{EIG3_MT}. The fit of Curie-Weiss expression $\chi(T)~= \mu^2_{eff}/(8(T - \theta_{\rm p}))$ to $\chi^{-1}$ between 100 and 300 K provides the following parameters: $\mu_{eff}$~=~7.94 and 7.65~$\mu_B$, and $\theta_{\rm p}$~=~-22.1 and -13.7~K for $H~\parallel$~[100] and [001], respectively. The value of $\mu_{eff}$ along [100] matches exactly with the Hund's rule derived value. An antiferromagnetic interaction between the divalent Eu ions is inferred from the negative values of $\theta_{\rm p}$ along the two directions. The susceptibility below 20~K is shown in the main panel of Fig.~\ref{EIG3_MT}. There is a cusp at $T_{\rm N1}$~=~12.4~K typical of antiferromagnetic ordering, followed by a mild shoulder at 7.3~K for $H~\parallel$~[001].

%
%
%

 On the other hand for $H~\parallel$~[100] there is a very subtle change of slope observed at 12.4~K and a very clear cusp is observed at 7.3~K. The main features of the susceptibility in the magnetically ordered state are  in good agreement with those reported in ref.~\onlinecite{Oleksandr_EuTGe3}. For a collinear bipartite antiferromagnet, the susceptibility below $T_{\rm N}$ is temperature independent along the magnetic hard axis while it gradually decreases to zero along the easy axis as the temperature is lowered to zero. In the present case the $\chi$ decreases along both [001] and [100] indicating a magnetic configuration which is different from simple collinear bipartite antiferromagnetic. The $T_{\rm N1}$ is comparable to $\theta_{\rm p}$ for $H~\parallel$~[001] but it is smaller in relation to $\theta_{\rm p}$ for $H~\parallel$~[100]. The latter may be due to the existence of ferromagnetic exchange interaction between the Eu ions at second near neighbour distance along [100], indicating a relatively complex magnetic structure.  It may be mentioned here that in ref.~\onlinecite{Oleksandr_EuTGe3} a single $\theta_{\rm p}$ value of -17~K is reported. 

\begin{figure*}[h]
\centering
\includegraphics[width=0.95\textwidth]{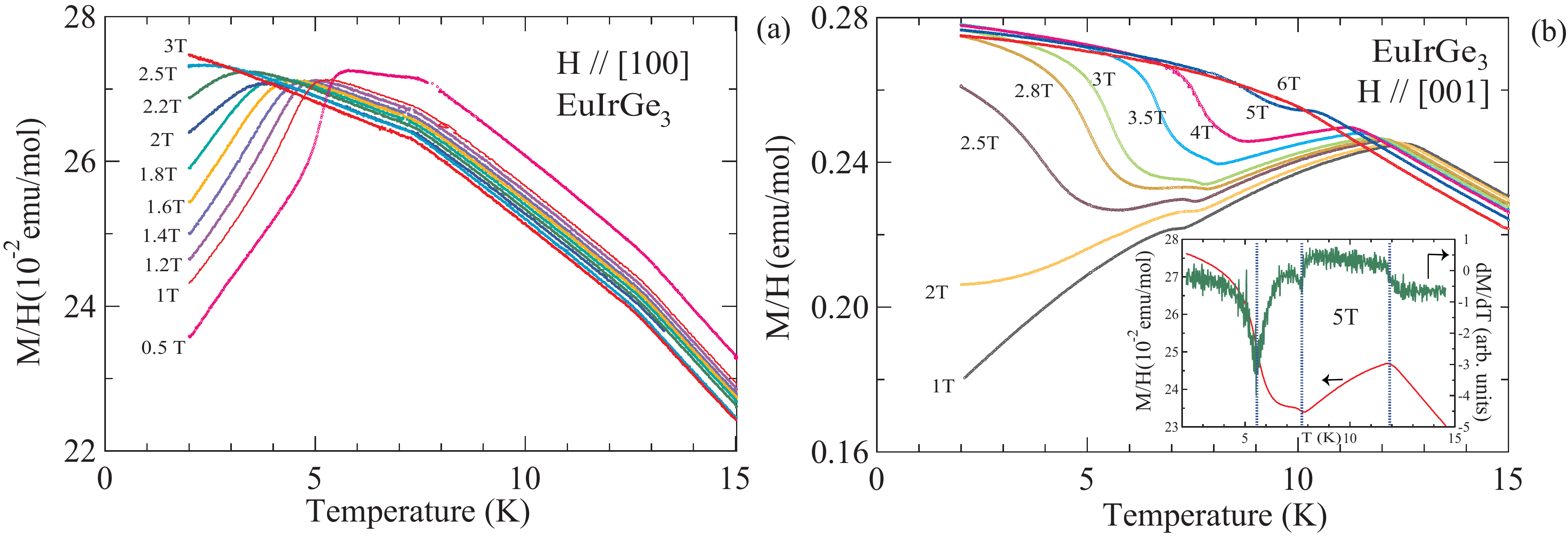}
\caption{\label{Chi_T_H_EIG3} Field dependence of $M/H(T)$ along [100] (left panel) and [001] (right panel) for EuIrGe$_3$. Inset in  (b) shows $M/H(T)$ at 5~T on the left scale and its derivative on the right scale with vertical dotted line at critical points.}
\end{figure*}

	The magnetisation, $M/H$, along [100] and [001] was measured in various fields ranging from 0.5 to 6~T and the data below 15~K are shown in Fig.~\ref{Chi_T_H_EIG3}. Additional features appear as the field is increased above 0.1~T.
	%
%
	 At 0.5~T, $M/H$ ($H~\parallel$~[100]) shows a knee near 5.5~K, which shifts to lower temperatures as the field is increased and either vanishes at H~=~3~T or occurs below 1.8~K. The anomalies at 12.4~K and 7.3~K also shift slightly to lower temperatures with increasing field. For $H~\parallel$~[001], the magnetisation between 2.5 and 4~T shows a prominent upturn at low temperatures, indicating a field-induced change in the direction of the magnetic moments. The mild shoulder at 7.3~K shifts to higher temperatures with increasing field while the peak at 12.4~K (in 0.1~T field) shifts to lower temperatures. The two appear to merge in a field of 6~T. At high fields, $\sim$3~T and above, the magnetisation $M/H$ at temperatures approaching 1.8~K is comparable for both the directions. These plots show that the configuration of the Eu moments in the magnetically ordered regime is modified by the applied field. The critical points derived from $M/H(T,H)$ have been included while constructing $H-T$ phase diagram in Fig.~\ref{Phase_Diagram}. 
	
\begin{figure}[h]
\centering
\includegraphics[width=0.50\textwidth]{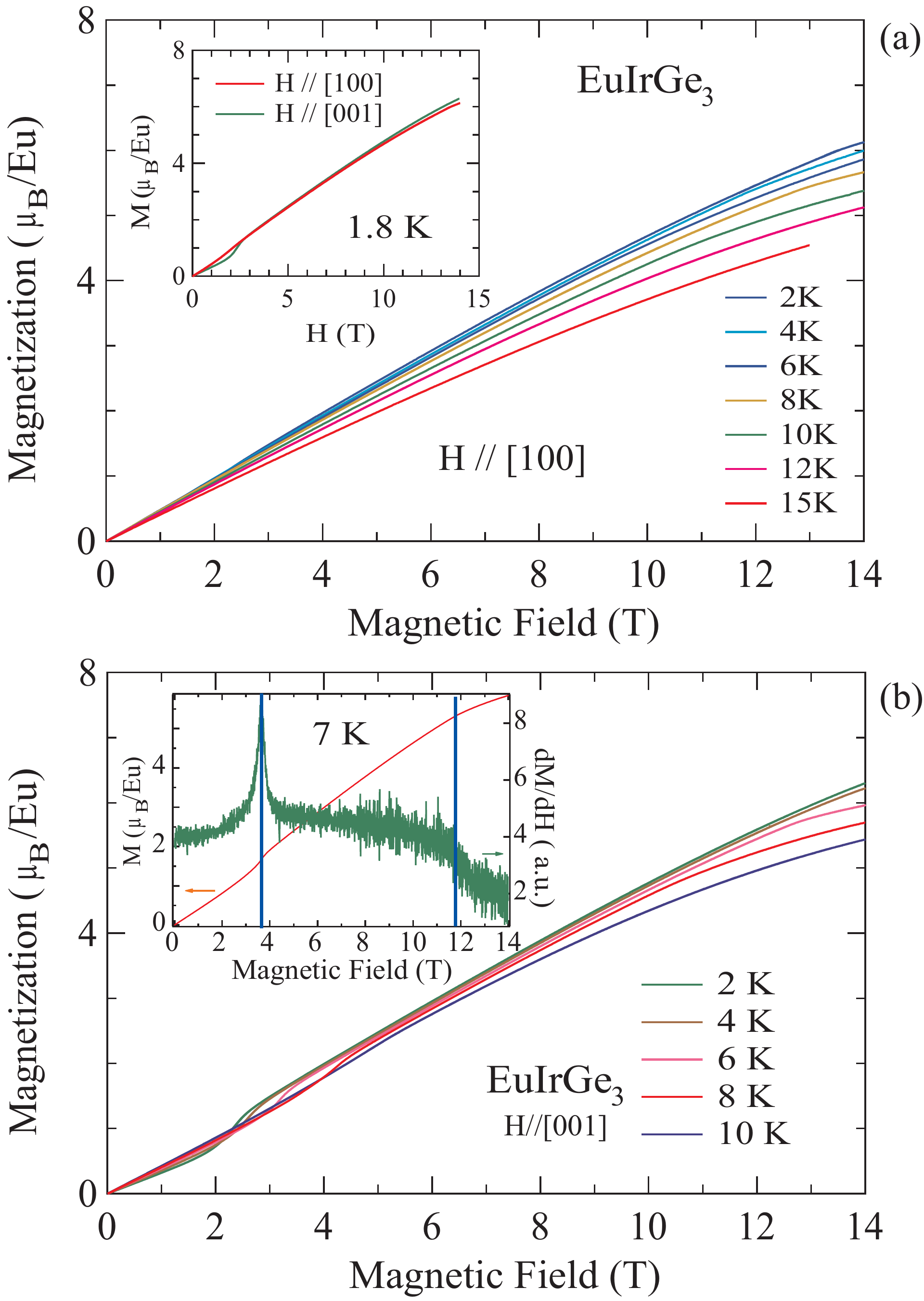}
\caption{\label{MH_EIG3} (a) Temperature dependence of $M(H)$ in EuIrGe$_3$ along (a) [100] and (b) [001]. Inset in (a) represents nearly isotropic magnetisation plots in EuIrGe$_3$ and in (b) a representative derivative of magnetisation plot along [001] at 7~K (right scale) along with the $M(H)$ data (left scale) are shown.}
\end{figure}

	The magnetisation at selected temperatures for $H~\parallel$~[100] and [001] in applied fields up to 14~T is shown in Figs.~\ref{MH_EIG3}a and~\ref{MH_EIG3}b, respectively. In the inset of Fig.~\ref{MH_EIG3}a the data at 1.8~K along the two directions are plotted. Above 3~T and between 3 and 8~T, the magnetisations at 1.8~K along the two directions virtually overlap (in conformity with the $M-T$ data discussed above) and then slightly bifurcate at higher fields. A spin-flop like behaviour is clearly seen near 2~T for $H~\parallel$~[001] (which shifts upward in field as the temperature is increased (see, Fig.~\ref{MH_EIG3}b) and eventually vanishes at 8~K, while it is barely discernible along [100] occurring at $\sim$1.5~T. 
%
%
	
	Though the variation of the magnetisation with field and temperature is consistent with the antiferromagnetic nature of the magnetic transition in EuIrGe$_3$, it is not possible to infer the easy and the hard axis of magnetisation. For a collinear two-sublattice antiferromagnet, the magnetisation should be zero below the spin-flop field along the easy axis, but such a behaviour is not seen. At 1.8~K the magnetisation at 14~T (6.3~$\mu_B$/Eu) along [001] is slightly larger than (6.15~$\mu_B$/Eu) along [100] and the saturation is not yet reached along both the directions.  It appears that the spin-flip field at 1.8~K is nearly independent of the direction like in EuPtGe$_3$, indicating a small anisotropy. 

\begin{figure}[h]
\centering
\includegraphics[width=0.50\textwidth]{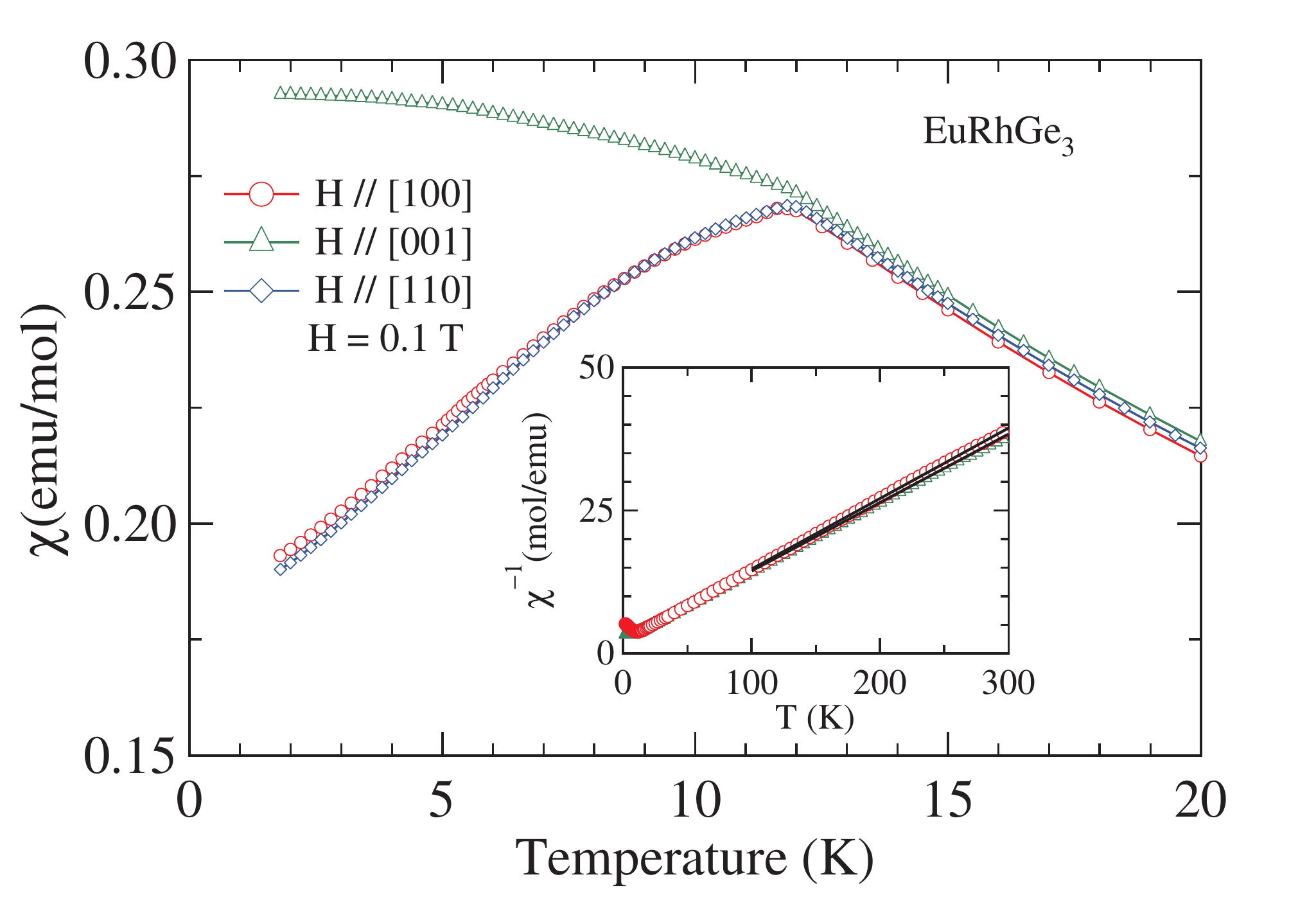}
\caption{\label{ERG3_MT} Anisotropic magnetic susceptibility and inverse magnetic susceptibility (in inset) as a function of temperature at magnetic field 0.1~T of EuRhGe$_3$.}
\end{figure}

	In contrast to EuIrGe$_3$, the Rh analog shows only one antiferromagnetic transition which occurs close to $T_{\rm N}$~=~12~K in applied field of 0.1~T (see, Fig.~\ref{ERG3_MT}), in accordance with Ref.~\onlinecite{Oleksandr_EuTGe3}. Fig.~\ref{ERG3_MT} shows the data below 20~K for $H~\parallel$~[100], [110] and [001], respectively. It is noticed that the response in the $ab$-plane is isotropic. Above 100~K, the inverse susceptibility fits well to the Curie-Weiss law with  the values: $\mu_{eff}$~=~7.56 and 7.78~$\mu_B$, $\theta_{\rm p}$~=~-7 and -11~K for $H~\parallel$~[100] and [001] respectively. It is likely that the value of  $\mu_{eff}$ is slightly lower due to the presence of tiny inclusions of In metal in the crystal which get incorporated during the crystal growth. As a result the amount of Eu used in the calculation of  $\mu_{eff}$ is actually slightly overestimated. We infer the presence of In from the slight drop in the susceptibility measured in an applied field of 0.005~T near the superconducting transition temperature of In. The drop vanishes when the applied field is increased to 0.05~T. From the susceptibility plots below $T_{\rm N}$ one can infer that the hard axis of magnetisation is aligned close to [001] direction.  
%
%
	
%
\begin{figure*}[!]
\centering
\includegraphics[width=0.95\textwidth]{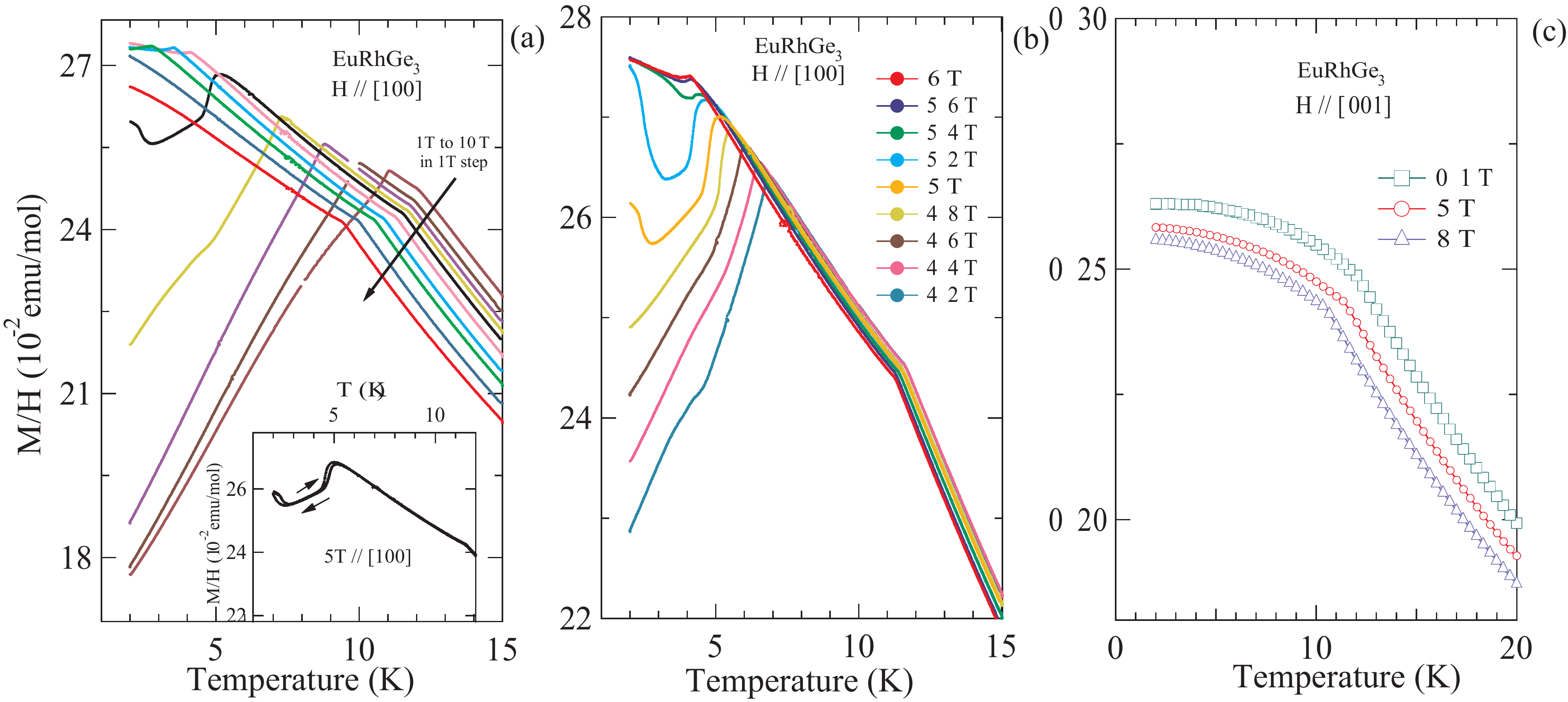}
\caption{\label{Chi_T_H_ERG3} Magnetisation $M/H(T)$ at various fields along [100] (a-b) and [001] (c) for EuRhGe$_3$. Inset in (a) shows the hysteresis in the data taken in the warming and cooling cycle at 5~T.}
\end{figure*}
%
		 
	At higher fields applied parallel to [100] the transition shifts to lower temperatures reaching 9.4~K in 10~T (see, Fig.~\ref{Chi_T_H_ERG3}a). An additional peak which occurs close to 11~K in 1~T gradually shifts to 2.8~K in 8~T. Between 4 and 6~T, the magnetisation shows additional features below 5~K which are depicted in Fig.~\ref{Chi_T_H_ERG3}b. The field induced changes in the magnetic configuration have a first order character as indicated by the hysteresis observed at 5~T (see inset of Fig.~\ref{Chi_T_H_ERG3}a). The data depicted in Figs.~\ref{Chi_T_H_ERG3}a and \ref{Chi_T_H_ERG3}b  indicate a field induced complex phase diagram. On the other hand, for fields along the [001] direction the character of the plots remains unaltered except  that $T_{\rm N}$ decreases with field (see Fig.~\ref{Chi_T_H_ERG3}c). While our observations do not provide us the actual configuration of the magnetic moments, they appear to suggest a non collinear antiferromagnetic structure which evolves in a complex fashion for field applied in the $ab$-plane.
%
\begin{figure}[!]
\centering
\includegraphics[width=0.50\textwidth]{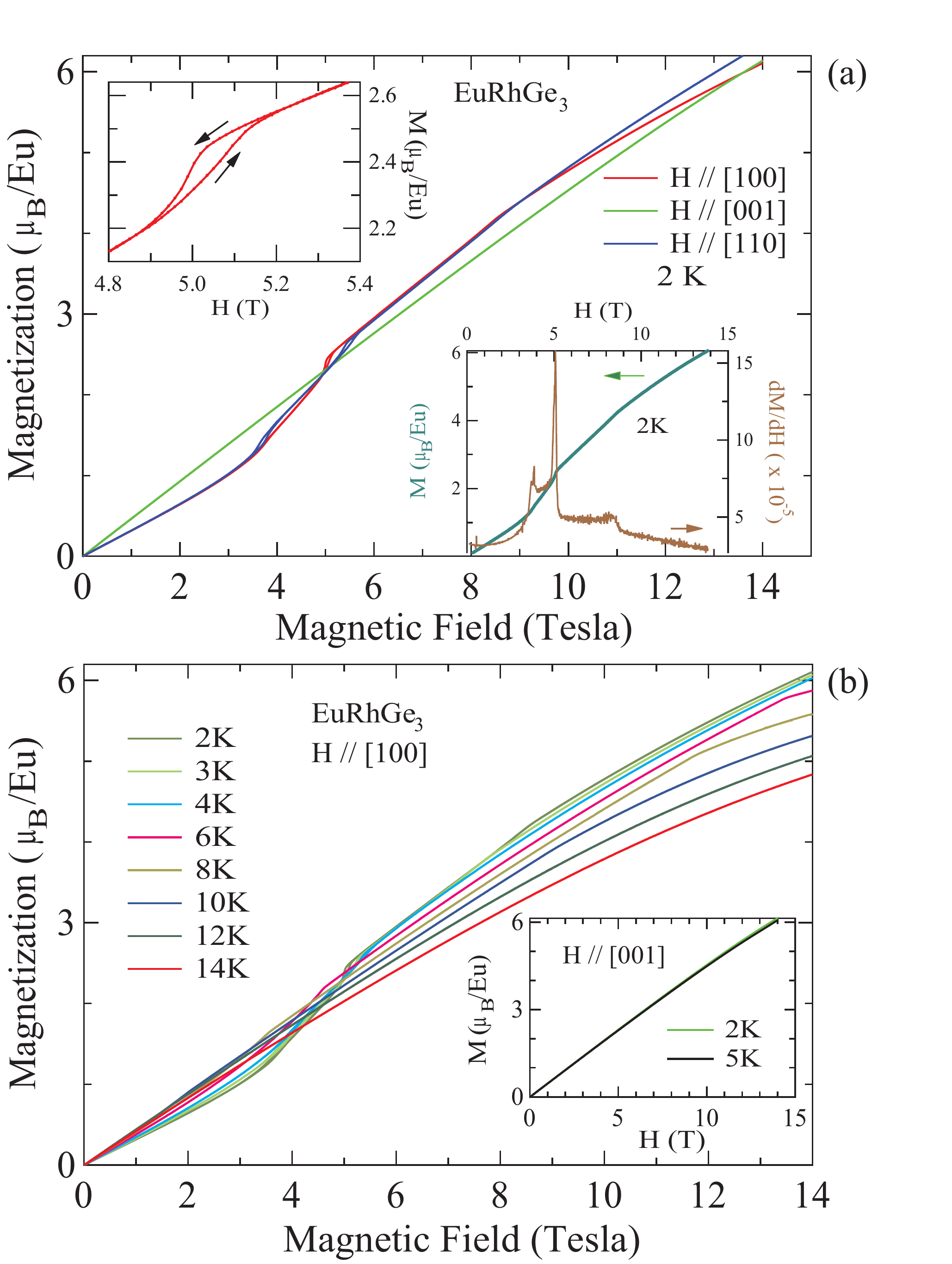}
\caption{\label{MH_ERG3} (a) Isothermal magnetisation $M(H)$ at 2~K of EuRhGe$_3$ along the principal crystallographic directions. Bottom inset shows $M(H)$ at 2~K on left scale and its derivative on right scale revealing the field induced spin reorientation. (b) Temperature evolution of $M(H)$ of EuRhGe$_3$ along [100] in main panel and along [001] in the inset.}
\end{figure}
%
 	
	The main panel of Fig.~\ref{MH_ERG3}a shows the isothermal magnetisation plots at 2~K for $H~\parallel$~[100], [110] and [001]. The magnetisation along [100] at various temperatures is plotted in Fig.~\ref{MH_ERG3}b; the inset shows the magnetisation at 2 and 5~K for $H~\parallel$~[001].
	
	 Along [001] the magnetisation increases nearly linearly with the field reaching a value of 6.2~$\mu_B$/Eu ion at 14~T. There is hardly any change in the magnetisation at 2 and 5~K(Fig.~\ref{MH_ERG3}b inset). On the other hand one sees additional features for $H~\parallel$~[100] and [110] with evidence of spin-flop transitions in 3-4~T range and near 5~T. Above 10~T the $ab$-plane isotropy of the magnetisation is slightly violated. At the highest field (14~T) the magnetisations along the three directions are nearly identical. A derivative plot of magnetisation for $H~\parallel$~[100], $dM/dH$, shown in the inset of Fig.~\ref{MH_ERG3}a shows three peaks which are a signature of the field induced changes in the magnetic configuration. The magnetisation shows hysteresis around 5~T at 2~K (see, upper inset Fig.~\ref{MH_ERG3}a) which correlates nicely with the data depicted in the inset of Fig.~\ref{Chi_T_H_ERG3}a. Temperature dependence of $M(H)$ along [100] is shown in main panel of Fig.~\ref{MH_ERG3}b. 
	 
\subsection{Heat capacity}

\begin{figure}[!]
\centering
\includegraphics[width=0.6\textwidth]{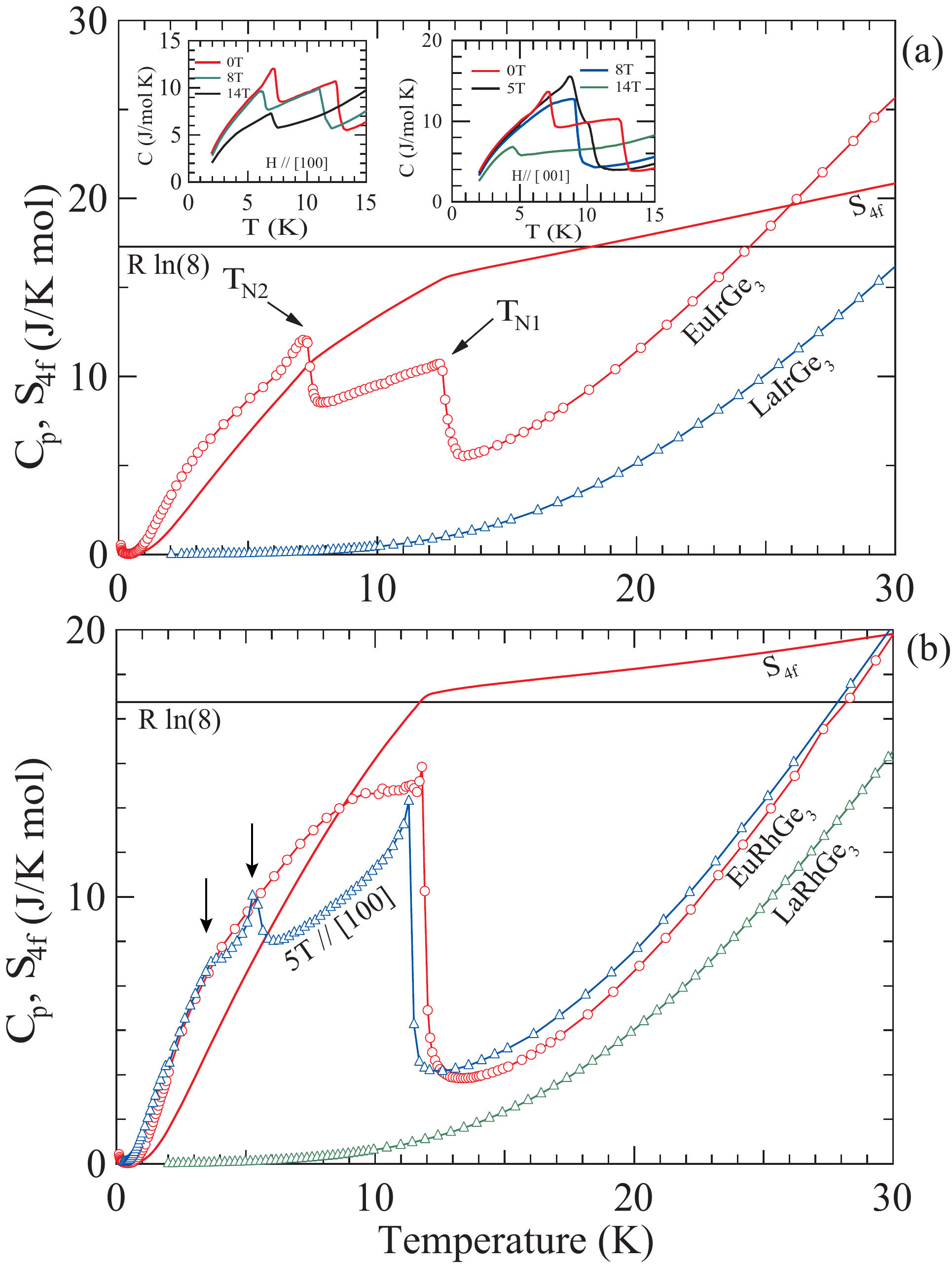}
\caption{\label{HC} Heat capacity ($C_p(T)$ and calculated magnetic entropy ($S_{\rm 4f}$)  of (a) EuIrGe$_3$ and (b) EuRhGe$_3$, respectively. $C_p(T)$ of corresponding La-analogues are also shown in the main panels. Insets in (a) show the $C_p(T)$ curves when field is applied along [100] (left) and [001] (right). In (b), blue trace represents data taken at 5~T field applied parallel to $a$-axis, capturing the field induced phase transitions indicated by arrows. }
\end{figure}

The specific heat measured down to 100~mK in zero field (Fig.~\ref{HC}a) confirms the presence of two transitions in EuIrGe$_3$, at $T_{\rm N1}$~=~12.4 and $T_{\rm N2}$~=~7.2~K, in close correspondence with the low-field magnetisation data presented above. The magnitude of the jump in the heat capacity at $T_{\rm N1}$, $\sim$5~J/mol K, which is far below the mean-field value of 20.14~J/mol~K for a mol of spin $S$~=~7/2. This suggests that the transition at $T_{\rm N1}$ is from paramagnetic to amplitude modulated antiferromagnetic configuration. At $T_{\rm N2}$ the transition from this intermediate state to an equal moment antiferromagnetic configuration takes place, as confirmed by $^{151}$Eu M\"{o}ssbauer spectra (see below). 

The heat capacity was also measured in applied fields of 8 and 14~T with $H~\parallel$~[100]. At 8~T, the two peaks at 7.2 and 12.4~K (zero field) have shifted slightly lower in temperature to 6.1 and 11~K, respectively, in correspondence with the magnetisation data shown in Fig.~\ref{Chi_T_H_EIG3}. The jump in the heat capacity is slightly reduced.  At 14~T, there is only one peak at 7~K, with appreciable reduction in peak height. For $H~\parallel$~[001], at 5~T both peaks come closer but are still well resolved; at higher fields we observe only a single peak in agreement with magnetic phase diagram in Fig~\ref{Phase_Diagram}b.
%
%

	The heat capacity of the iso-structural LaIrGe$_3$ is also plotted in Fig.~\ref{HC}a, and the $4f$ contribution to the heat capacity, $C_{4f}$, and entropy $S_{4f}$ were calculated under the assumption that the phonon heat capacities of LaIrGe$_3$ and EuIrGe$_3$ are identical, after normalization due to the slightly different atomic masses of La and Eu. The entropy attains the value of $Rln8$ (for Eu$^{2+}$ ions, $S$~=~7/2 and $L$ =0) near 18~K but keeps on increasing at higher temperatures, indicating a poor validity of the assumption of identical phonon spectra in LaIrGe$_3$ and EuIrGe$_3$ at least at higher temperatures.  A similar situation was earlier encountered in EuPtSi$_3$~\cite{Neeraj_EuPtSi3}. The $C/T$ vs. $T^2$ plot of LaIrGe$_3$ is linear below 8~K, characterized by $\gamma$~=~4.0 mJ/mol K$^2$ and $\beta$~=~0.349 mJ/mol K$^4$. A Debye temperature, $\theta_{\rm D}$ of 303~K is inferred from $\beta$. 
	
	The heat capacity data of EuRhGe$_3$ measured in zero and 8~T ($H~\parallel$~[100]) are plotted in Fig.~\ref{HC}b. The relatively sharp jump in the heat capacity near 12~K in zero field is in excellent agreement with the magnetisation data discussed above. Surprisingly, the jump in the heat capacity at $T_{\rm N}$ in EuRhGe$_3$ is about 13~J/mol K, which is far below the value for a transition to an equal moment antiferromagnetic state (20.14~J/mol~K) in the mean field model~\cite{Blanco}. It may be noted that the shape of the heat capacity variation below T$_N$ is rather unusual and  similar to some cases of amplitude modulated moment state described in Fig.5 in ref.~\onlinecite{Blanco}. In applied fields the magnetic transition shifts to lower temperatures and additional peaks, marked by downward arrows in Fig.~\ref{HC} for 5~T data are observed, in conformity with the in-field magnetisation data (Fig.~\ref{Chi_T_H_ERG3}) described above.  The heat capacity of non-magnetic, reference analogue LaRhGe$_3$ is also plotted. The entropy $S_{\rm 4f}$ was estimated using the method mentioned above. It again exceeds the maximum value of \textit{Rln8} like in the Ir compound. For LaRhGe$_3$, $\gamma$~=~6.7 mJ/mol~K$^2$ and $\beta$~=~0.376~mJ/mol K$^4$. A Debye temperature, $\theta_{\rm D}$ of 296~K is inferred from $\beta$, which is close to that of the Ir compound.  
\subsection{Electrical Resistivity} 
\begin{figure*}[!]
\centering
\includegraphics[width=0.95\textwidth]{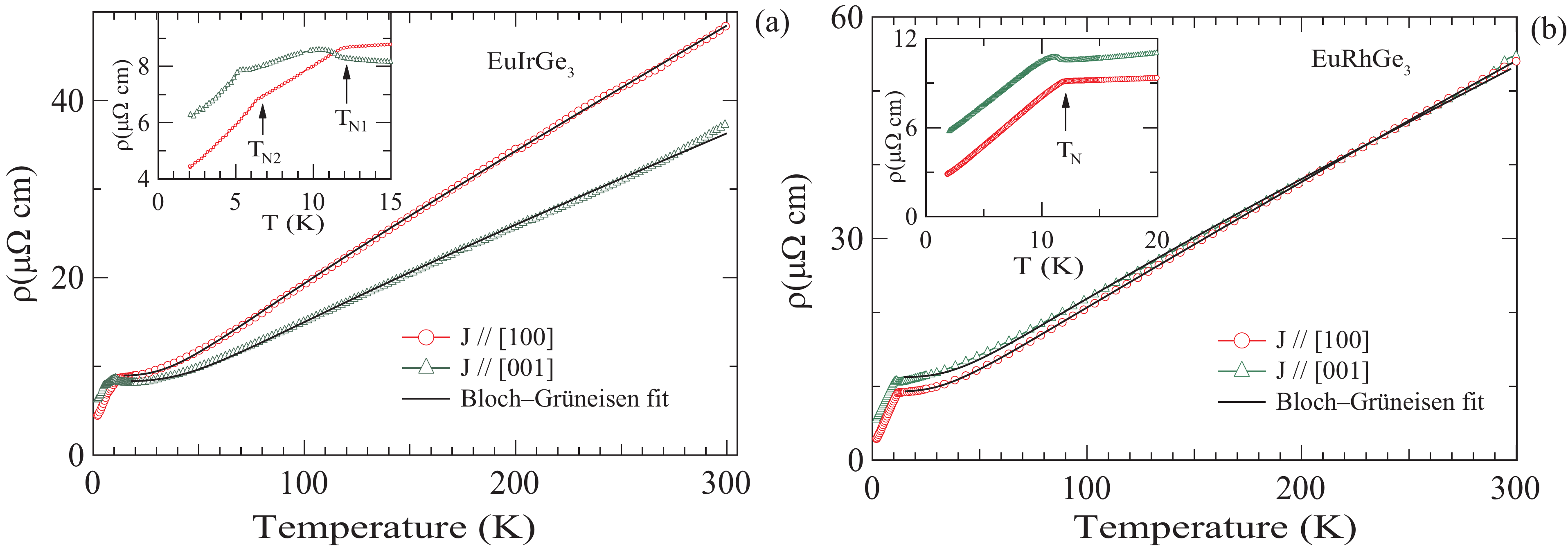}
\caption{\label{RT_1} Variation of electrical resistivity ($\rho(T)$ for current density $J~\parallel$~[100] and [001] for (a) EuIrGe$_3$ and (b) EuRhGe$_3$. Insets show low temperature data on an expanded scale where the magnetic transitions are marked by arrows.}
\end{figure*}

	The electrical resistivity of EuIrGe$_3$ and EuRhGe$_3$ with the current density $J$ parallel to [100] and [001], respectively is shown in Fig.~\ref{RT_1}. The resistivity shows anomalies for EuIrGe$_3$ at the two transitions $T_{\rm N1}$ and $T_{\rm N2}$ along both directions though with slightly different characteristics. Along [100] the resistivity decreases faster at each transition due to the rapid reduction in of the spin-disorder scattering. Above $T_{\rm N1}$ the resistivity monotonically increases up to the room temperature. On the other hand, for $J~\parallel$~[001]  at $T_{\rm N1}$ there is a slight upturn on cooling. The upturn at $T_{\rm N1}$ is suggestive of a gap-opening at the Fermi surface along [001] direction with AFM order, often referred as superzone gap. Many other rare earth intermetallics have been found to show this kind of behaviour.\cite{Pranab_CeGe, Anand_EuPd2As2, Hossain}
	
	 The electrical resistivity of EuRhGe$_3$ has some similarities with that of EuIrGe$_3$ described above. For $J~\parallel$~[100] the resistivity decreases at the single antiferromagnetic transition, while for $J~\parallel$~[001] the initial increase at $T_{\rm N}$~(=~12~K) again indicates the opening of a superzone gap like in EuIrGe$_3$. 

	We have fitted the Bloch Gr\"{u}neisen expression to our $\rho(T)$ data in the paramagnetic region given by following expression:
\begin{equation}
\label{BG Equation}
  \rho(T) = A+B\left(\frac{T}{\theta_{\rm R}}\right)\int_0^{\theta_{\rm R}/T} \frac{x^5}{\left(e^x-1\right)\left(1-e^{-x}\right)}dx
\end{equation}
where $\theta_{\rm R}$ is the Debye temperature determined from the $\rho(T)$ data, $A$ is the temperature independent part of resistivity comprising of electron scattering caused by crystal imperfection and spin disorder in paramagnetic state. $B$ is a material dependent prefactor. Parameters determined from the fit are listed in Table~\ref{Table_BG_fit}. It may be noted that the magnitude of $\theta_{\rm R}$ is different from $\theta_{\rm D}$. This is not unusual as $\theta_{\rm R}$ considers only the longitudinal lattice vibrations.

\begin{table}[h]
\centering
\caption{\label{Table_BG_fit} Parameters derived from the Bloch Gr\"{u}neisen fit to the $\rho(T)$ data of EuRhGe$_3$ and EuIrGe$_3$.}
\begin{tabular}{ccccc}
\hline \hline
 &\multicolumn{2}{c}{EuRhGe$_3$} & \multicolumn{2}{c}{EuIrGe$_3$} \\ \hline
 & $A(\mu\Omega$ cm) & $\theta_{\rm R}\rm(K)$ &  $A(\mu\Omega$ cm) & $\theta_{\rm R}\rm(K)$   \\ \hline
$J~\parallel$~[100] & $9.5$ & $246$ & $8.9$ & $224$ \\
$J~\parallel$~[001] & $11.2$ & $244$ & $8.3$ & $270$ \\
\hline
\end{tabular}
\end{table}
%
%
%

\begin{figure*}[!]
\centering
\includegraphics[width=0.95\textwidth]{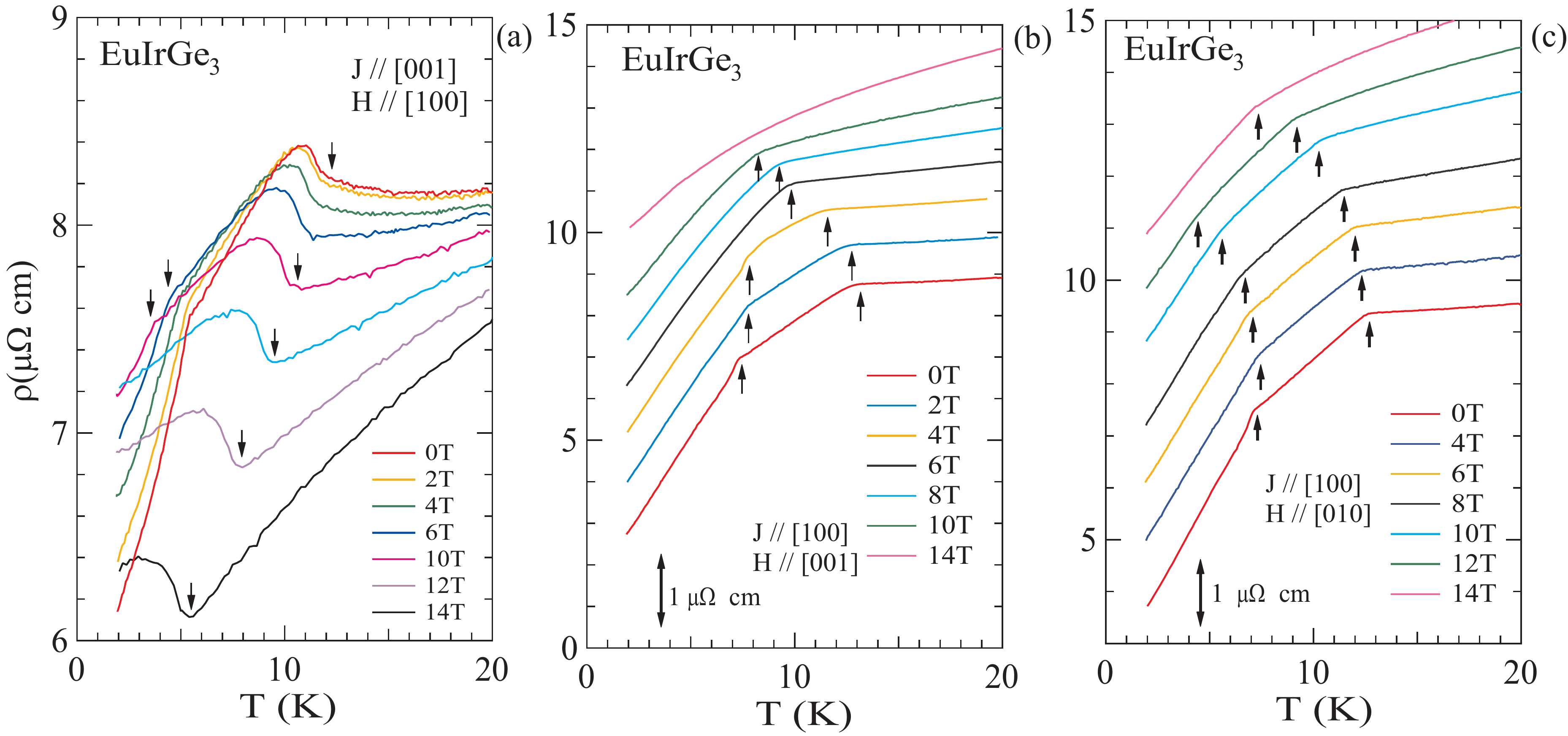}
\caption{\label{EIG3_RT_H} $\rho(T)$ of EuIrGe$_3$ for different configurations of current and applied magnetic fields. In b and c, the traces for non zero fields have been moved in vertical direction for clarity. Arrows indicate magnetic phase transitions.}
\end{figure*}

\begin{figure*}[!]
\centering
\includegraphics[width=0.95\textwidth]{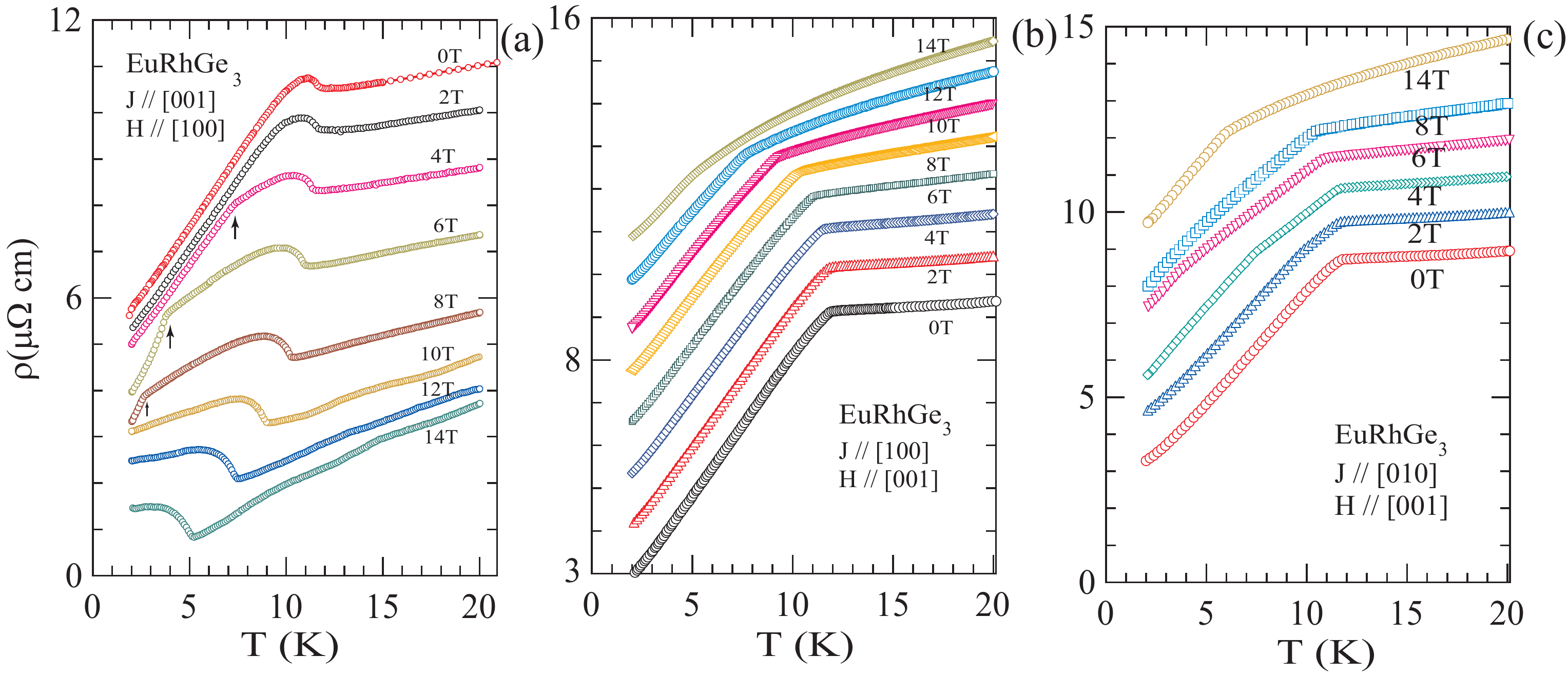}
\caption{\label{ERG3_RT_H} $\rho(T)$ of EuRhGe$_3$ for different configurations of current and applied magnetic fields. $\rho(T)$ curves other than 0~T have been shifted vertically for clarity.}
\end{figure*}

The transverse magnetic field dependence of electrical resistivity under different configurations is shown in Fig.~\ref{EIG3_RT_H} and ~\ref{ERG3_RT_H} for EuIrGe$_3$ and EuRhGe$_3$, respectively. The main features in the $\rho(T)$ data of both compounds are in excellent correspondence with the magnetic susceptibility data. The upturn in the resistivity at T$_N$ becomes more prominent as the field is increased (see, Fig. 9a and 10a), suggesting an enhancement of the superzone gap in the two compounds. 
	
%
%
	
\begin{figure*}[!]
\centering
\includegraphics[width=0.99\textwidth]{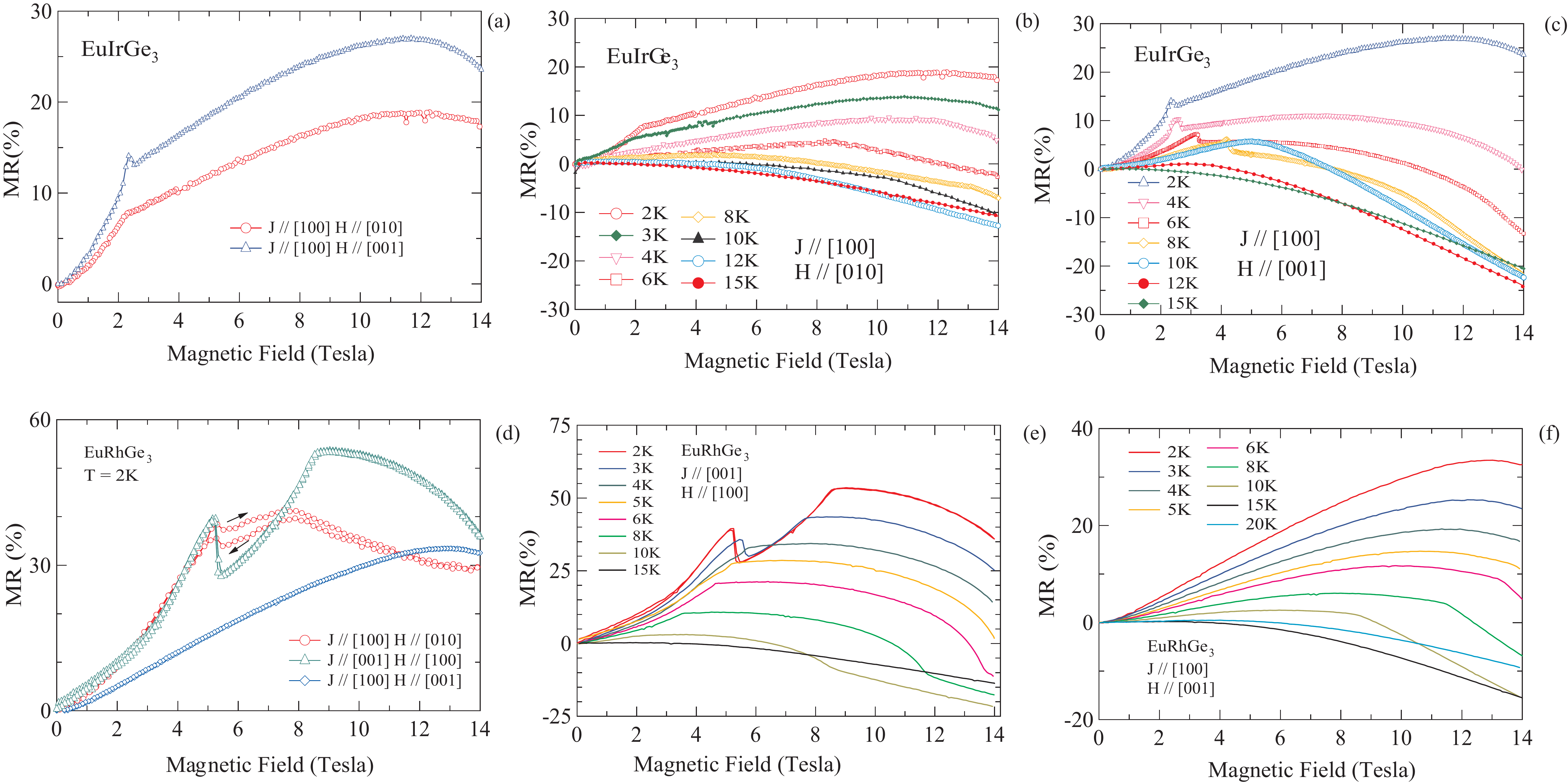}
\caption{\label{MR} The variation of the isothermal magnetoresistance $MR(H)$ with temperature in EuIrGe$_3$ (upper panels) and  EuRhGe$_3$ (lower panels).}
\end{figure*}

	The magnetoresistivity $MR$, defined as $MR(H)=(\rho(H)-\rho(H=0))\times100/\rho(H=0)$ of EuIrGe$_3$ for diffrent transverse configurations is shown in Figs.~\ref{MR}(a-c). For $H~\parallel$~[001], [010] and $J~\parallel$~[100] at T~=~2~K, the $MR$ is positive, increases rapidly with field and shows a minor anomaly near 2~T which corresponds well with the spin-flop transition seen in the magnetisation. The positive $MR$ is typically expected in an antiferromagnet as the field disrupts the antiferromagnetic ordered state. The positive $MR$ peaks near 12~T and then decreases slightly, indicating the proximity of the spin-flip field around 14~T. As the temperature is raised (see, Fig. 10c) the minor anomaly shifts to higher fields tracking the corresponding increase in the spin-flop transition field in the magnetisation, and the magnitude of positive $MR$ decreases due to the increase of temperature.  At 6~K the $MR$ becomes negative for $H$ $>$ 10~T, and the crossover field value decreases with further increase of temperature. At 15~K the $MR$ in the paramagnetic state is negative at all fields, most likely due to the ordering effect of the field on the fluctuating moments.

%
%
	
The $MR$ for $H~\parallel$~[010] shows qualitatively similar field dependence (see, FIg. 10b) as described above for $H~\parallel$~[001]. The anomalies observed in $MR$ data are included in phase diagram corresponding well to the magnetisation data. 
	   
%
%
	The $MR$ of EuRhGe$_3$ at selected temperatures is shown in Figs~\ref{MR}(d-f). Fig.~\ref{MR}d shows a comparison between $MR$ data taken at 2~K in different configurations. The nature of $MR$ curves is dependent upon the field direction as well as the direction of current density $J$. For $H~\parallel$~[100] and $J~\parallel$~[001] the $MR$ at 2~K is positive and initially increases with field. It shows anomalies at ~5 and ~8.5 T which mirror the spin-flop transitions seen in the magnetisation at these fields (Fig.\ref{Chi_T_H_ERG3}b). Above 8.5~T the $MR$ is still positive but begins to decline in its absolute values most likely due to the increasing alignment of the moments along the field direction as the spin-flip field is approached.  As the temperature is increased the two anomalies approach each other, shifting in opposite directions, and above $\sim$~4-5 K they apparently merge and then the single anomaly shifts to lower fields with increasing temperature (see, Fig. 11e). At 6~K, the $MR$ shows an anomaly in 13-14~T range which shifts to 11-12 and 8-9~T intervals at 8 and 10~K, respectively. This feature matches well with the magnetisation plots measured at these temperatures (indicated by arrows in Fig.~\ref{MH_ERG3}b).  In the paramagnetic region (15 and 20~K) the $MR$ is negative. Lastly, the $MR$ data for $H~\parallel$~[001] and $J~\parallel$~[100] are shown in Fig.~\ref{MR}f. It may be recalled that [001] is relatively the hard-axis of magnetisation in EuRhGe$_3$. The $MR$ up to 5~K is positive, increasing with field and showing a slight decline above 12~T. At 8 and 10~K the decline in $MR$ is marked by a sharp knee at $\sim$11.5 and $\sim$9~T, respectively, which are phenomenologically similar to the one's seen in Fig.~\ref{MR}e and occur at similar values of fields as well, and may have a similar origin. Again, the $MR$ in the  paramagnetic region is negative. 
\subsection{Magnetic phase diagram}
\begin{figure*}[h]
\centering
\includegraphics[width=0.95\textwidth]{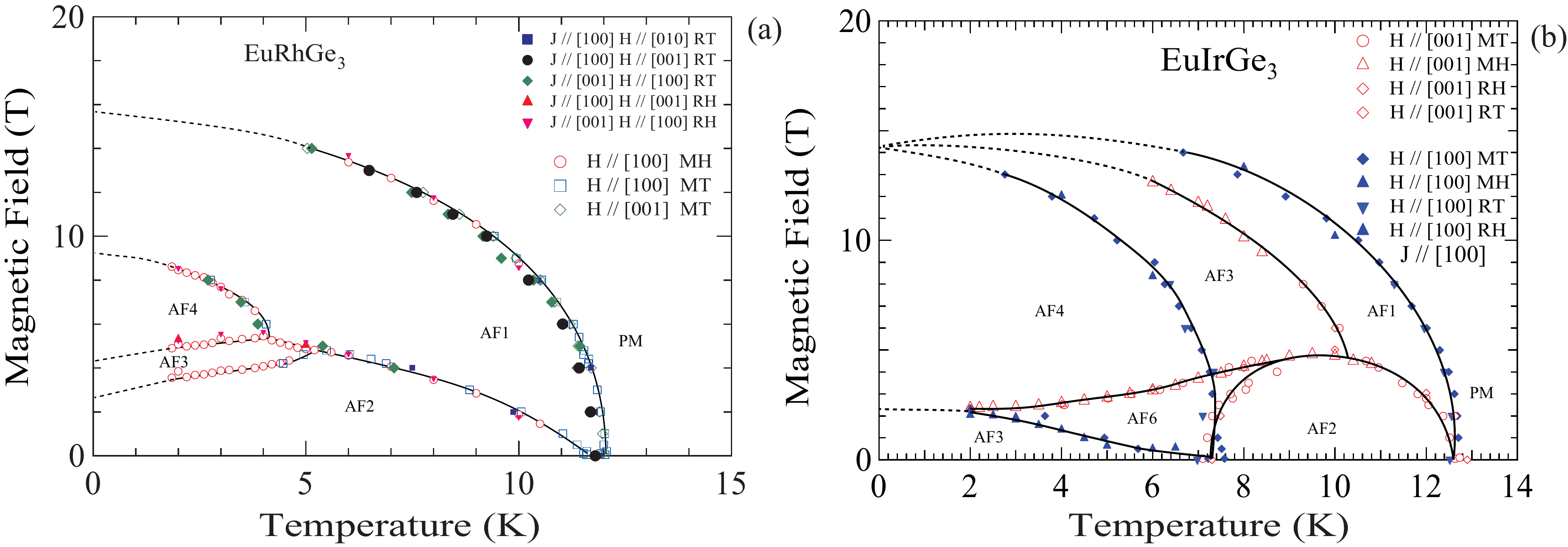}
\caption{\label{Phase_Diagram} $H-T$ phase diagram of (a) EuRhGe$_3$ and (b) EuIrGe$_3$ constructed from critical points in $M(T,H)$ and $\rho (T,H)$ data. Solid lines are guide to the eye and dotted lines are extrapolation.}
\end{figure*}

From $M(T,H)$ and $\rho (T,H)$ data we have constructed the magnetic phase diagrams of EuRhGe$_3$ and EuIrGe$_3$ shown in Fig.~\ref{Phase_Diagram}.
The conclusions derived from these two sets of data correspond very well with each other. AF1, AF2,... denote  phases specified by different antiferromagnetic configurations. In EuRhGe$_3$, at low fields a second transition appears which shifts to lower temperatures with increasing magnetic field until the occurrence of a  tricitical point at (5~K, 4.9~T), followed by another tricritical point at (4~K, 5.4~T). Dotted lines are plausible extrapolations.
%
%
EuIrGe$_3$ magnetic phase diagram also shows a similar degree of complexity. Here the red symbols have been used for points determined from $H~\parallel$~[001] and blue symbols for $H~\parallel$~[100] data, respectively for the $M$ vs $T$, $M$ vs $H$, $R$ vs $H$ and $R$ vs $T$ experiments. For $H~\parallel$~[001], on increasing the field, $T_{\rm N1}$ and $T_{\rm N2}$ come closer and merge together accompanied with the appearance of other field induced transitions forming a closed dome centered around 10~K. Above 5~T we could observe only one transition along $c$-axis. On the other hand, for field parallel to $a$-axis $T_{\rm N1}$ and $T_{\rm N2}$ are suppressed, apparently merging at around 14.2~T at absolute zero. Interestingly, it looks that the high field phase for $H~\parallel$~[001] seems to converge at the same point at 0~K as the merging point of two phase lines for $H~\parallel$~[100]. An additional low field phase line nearly parallel to the temperature axis, having opposite curvatures for $a$ and $c$ axes but merging with each other at around 2~K are observed.
It may be noted that the phase boundary between PM and AF1 phase is also a demarcation line for the superzone gap in both, EuRhGe$_3$ and EuIrGe$_3$. 
\subsection{$^{151}$Eu M\"{o}ssbauer spectra} 
\begin{figure}[h]
\centering
\includegraphics[width=0.80\textwidth]{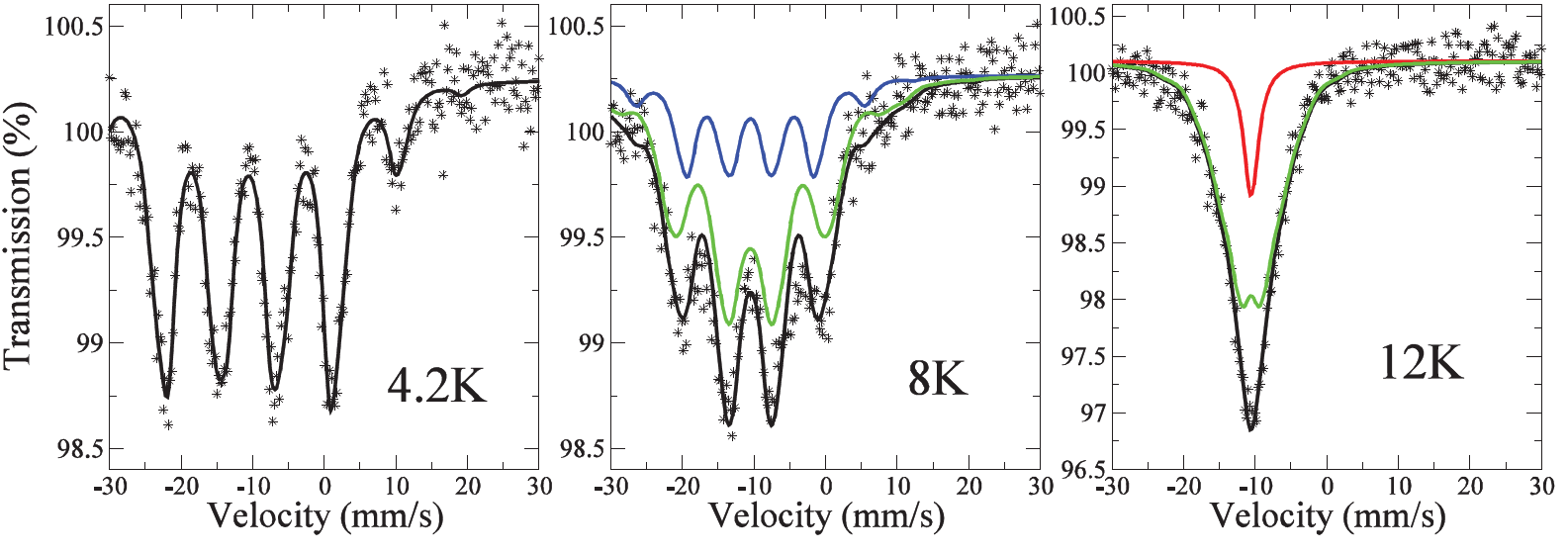}
\caption{\label{Mossbauer_EuIrGe3} $^{151}$Eu M\"{o}ssbauer spectra at selected temperatures in EuIrGe$_3$, in the equal moment phase (4.2~K), close to the equal moment \- incommensurate transition (8~K) and close to the incommensurate-paramagnetic transition (12~K). At 8~K and 12~K, the green subspectrum represents the incommensurate pattern, the blue subspectrum the equal moment  pattern and the red subspectrum the paramagnetic pattern..}
\end{figure}

The $^{151}$Eu M\"{o}ssbauer spectra at 4.2, 8 and 12~K in EuIrGe$_3$ are shown in Fig.~\ref{Mossbauer_EuIrGe3}. The 4.2~K spectrum is a regular hyperfine field pattern with $H_{hf}$~=~28.9(2)~T, indicating an equal moment magnetic ordering. At 8~K, the spectrum has to be fitted to a superposition of a commensurate and of an incommensurate pattern, and at 12~K, close to $T_{\rm N1}$, to a superposition of an incommensurate pattern and of a single line characteristic of the paramagnetic phase.

%
%
	At 8~K and 12~K, the fits shown are probably not the unique solution because the two subspectra are not resolved. So they must be considered as indicative. The presence of two subspectra is however needed because fits with an incommensurate pattern alone yield unphysical modulations. One can conclude that the two transitions, from the paramagnetic to the incommensurate amplitude modulated phase and from the amplitude modulated to the equal moment phase, present a high degree of first order.
	
\begin{figure}[h]
\centering
\includegraphics[width=0.80\textwidth]{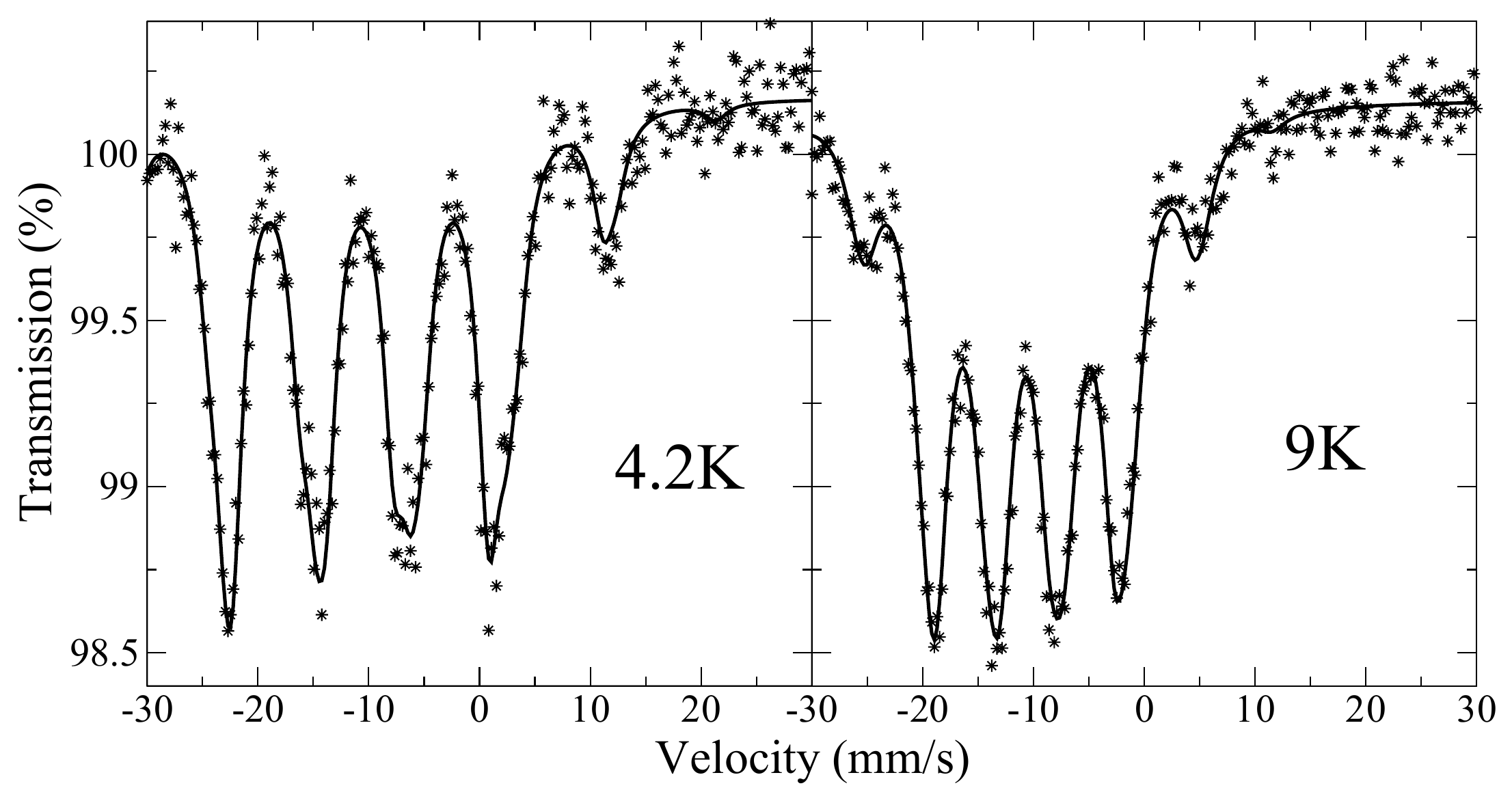}
\caption{\label{Mossbauer_EuRhGe3} $^{151}$Eu M\"{o}ssbauer spectra in EuRhGe$_3$ at 4.2~K and 9~K.}
\end{figure}

	For EuRhGe$_3$, the Mössbauer spectra recorded at 4.2 and 9~K, shown in Fig.~\ref{Mossbauer_EuRhGe3}, present evidence for a single magnetic transition. The $^{151}$Eu M\"{o}ssbauer spectrum at 4.2~K is standard with a hyperfine field of 29.8~T. 

%
\vspace{5 mm}
%
 
	Of the five compounds EuPtSi$_3$, EuPtGe$_3$, EuNiGe$_3$, EuIrGe$_3$ and EuRhGe$_3$ in which the $^{151}$Eu M\"{o}ssbauer data have been taken, with the exception of EuPtGe$_3$ and EuRhGe$_3$ the remaining compounds show a cascade of magnetic transitions, the intermediate phase being amplitude modulated. In EuPtGe$_3$, where the magnetisation is rather isotropic, no intermediate phase is present  and it was conjectured that multiple transitions may be linked to anisotropy~\cite{Arvind_EuNiGe3}. The isothermal magnetisation of EuIrGe$_3$ is rather similar to that of EuPtGe$_3$, yet it shows a cascade of transitions. This shows that  other factors  are important. Neutron diffraction studies are  required to find the antiferromagnetic configuration of the Eu moments in these compounds and its variation with temperature and field.

\end{document}